\documentclass[aps,prl,11pt,superscriptaddress]{revtex4}
\newcommand{\bcso}{BaCuSi$_2$O$_6$}
\newcommand{\scbo}{SrCu$_2$(BO$_3$)$_2$}
\newcommand{\hcl}{H_{c1}}
\newcommand{\hcu}{H_{c2}}
\usepackage[latin1]{inputenc}
\usepackage{graphicx,epsfig,epstopdf}
\begin{document}

\title{Bose-Einstein Condensation in Magnetic Insulators}
\author{Thierry Giamarchi}
\affiliation{DPMC-MaNEP, University of Geneva, 24 Quai Ernest Ansermet, 1211
Geneva 4, Switzerland}
\author{Christian R\"uegg}
\affiliation{London Centre for Nanotechnology and Department of Physics and
Astronomy, University College London, Gower Street, London WC1E 6BT, UK}
\author{Oleg Tchernyshyov}
\affiliation{Department of Physics and Astronomy, Johns Hopkins University,
Baltimore,MD 21218, USA}

\begin{abstract}
  {\bf The elementary excitations in antiferromagnets are magnons,
  quasiparticles with integer spin and Bose statistics. In an
  experiment their density is controlled efficiently by an applied
  magnetic field and can be made finite to cause the formation of a
  Bose-Einstein condensate (BEC). Studies of magnon condensation in a
  growing number of magnetic materials provide a unique window into an
  exciting world of quantum phase transitions (QPT) and exotic quantum
  states.}
\end{abstract}

\maketitle

\section{Introduction}

Not long after Bose and Einstein described quantum statistics of
photons \cite{Bose} and atoms \cite{Einstein}, Bloch applied the same
concept to excitations in solids.  He treated misaligned spins in a
ferromagnet as magnons, particles with integer spin and bosonic
statistics, to describe the reduction of spontaneous magnetisation by
thermal fluctuations \cite{Bloch}.  Matsubara and Matsuda pointed out
an exact correspondence between a quantum antiferromagnet and a
lattice Bose gas \cite{Matsubara}. It is thus natural to ask: can
these bosons undergo Bose-Einstein condensation and become superfluid?
The answer is yes. The question of order in spin systems and
possibility of BEC of magnons has been investigated theoretically for
a number of quantum antiferromagnets
\cite{batyev_spins_field,Affleck,Giamarchi,Nikuni,Matsumoto02,Matsumoto04,Rice,Nohadani,Wessel}.
Although in simple spin systems the concept of BEC can be applied, in
practice factors such as the large value of exchange constants and the
presence of anisotropies violating rotational symmetry may restrict
their usefulness. However, the analogy between spins and bosons has
shown to be very fruitful in those antiferromagnets where closely
spaced pairs of spins $S=1/2$ form dimers with a spin-singlet ($S=0$)
ground state and triplet ($S=1$) bosonic excitations called variously
magnons or triplons. In such systems BEC has been predicted to occur
\cite{Giamarchi} and was experimentally observed \cite{Nikuni,Ruegg03}
in the dimer system TlCuCl$_{3}$. This started a flourish of activity
on the subject and the hunt for such transitions in other magnetic
materials.

Other examples of condensates from elementary excitations were
recently found in magnetic thin films \cite{Demokritov} and
semiconductor microcavities \cite{Kasprzak}. Lattice bosons and
several transitions occurring in these systems have been studied
extensively on ultra-cold atomic gases in optical lattices
\cite{pitaevskii_becbook,Greiner,Bloch05,Bloch_RMP}. Here we present
an overview of recent developments in the field, and discuss how
quantum antiferromagnets offer several advantages that set them apart
from the other model systems in which the phenomenon of BEC occurs. We
also discuss ways to go beyond the physics of simple BEC condensation
and to look at the fascinating new quantum phases of interacting
bosons on a lattice.

\section{Bosons in magnets}
\label{sec:bosons}

Let us illustrate the basics of magnon BEC in real dimerized
antiferromagnets, such as extensively studied TlCuCl$_{3}$
\cite{Nikuni,Ruegg03,Cavadini01,Cavadini02,Matsumoto02,Matsumoto04,Johannsen,
 Vyaselev, Sherman,Misguich,Glazkov,Kolezhuk,Sirker} and \bcso\/
\cite{Sasago97,Ruegg07,Jaime,Sebastian,Batista,Vojta,Kramer}. The
lattice of magnetic ions in such materials (Fig. 1a) can be visualized
as a set of dimers, pairs of copper ions Cu$^{2+}$ carrying $S=1/2$
each and interacting via Heisenberg exchange:
\begin{equation}
\mathbf H = \sum_{i} J_{0} \, \mathbf S_{1,i} \cdot \mathbf S_{2,i}
+ \sum_{\langle mnij \rangle} J_{mnij} \, \mathbf S_{m,i} \cdot \mathbf S_{n,j}
- g \mu_B H \sum_{\langle ni \rangle} S^z_{m,i},
\label{eq:H}
\end{equation}
where $H$ denotes an external magnetic field in $z$--direction,
$i$,$j$ number dimers, and, $m,n=1,2$ their magnetic sites. The {\em
intra}-dimer exchange is the strongest interaction, which happens to
be antiferromagnetic, $J_0>0$, so that an isolated dimer has a ground
state with total spin $S=0$ and a triply degenerate excited state of
energy $J_0$ and spin $S=1$ (Fig. 1c).  It is convenient to identify
the triplet state with the presence of a triplon, a bosonic particle
with $S=1$, and the singlet state with the absence of a triplon, see
Fig. 1b.  As long as {\em inter}-dimer interactions are weak, the
ground state consists of non-magnetic singlets. It is disordered down
to absolute zero temperature without long-range magnetic order.  The
triplon excitations are made mobile by effective and weak {\em
  inter-}dimer couplings $J_{1,2,...} > 0$, from a sum over
single--ion interactions $J_{mnij}$, see Fig. 1b.  For dimers forming
a square lattice (for simplicity) the energy of a triplon with spin
projection $S^z = 0, \pm 1$ is
\begin{equation}
\varepsilon(\mathbf k) = J_0 + J_1[\cos{(k_x a)} + \cos{(k_y a)}]
- g \mu_B H S^z,
\label{eq:disp}
\end{equation}
where $\mathbf k = (k_x, k_y)$ is the particle wavevector, $a$ is the
lattice constant, and $D=4J_1$ the bandwidth, see Fig. 1c. The
energy-momentum dependence of the triplons and singlet-triplet
correlations have been measured directly by inelastic neutron
scattering \cite{Cavadini01,Sasago97,Ruegg07,Xu}.

The Zeeman term $- g \mu_B H S^z$ controls the density of triplons. As
the magnetic field increases, the excitation energy of triplons with
$S^z=+1$ is lowered and eventually crosses zero, as shown in Fig. 1c
and 2a.  This defines two critical magnetic fields $\hcl$ and $\hcu$
in the phase diagram, see Fig. 1d. At zero temperature, below $\hcl$
the magnetisation $m_z(H)$ is zero and only singlets exist. Between
$\hcl$ and $\hcu$ the magnetisation increases as the triplon band
fills up, see Fig. 1c.  Above $\hcu$ each site is occupied by a
triplon and the magnetisation saturates at one per dimer.

The bosonic nature of triplons is guaranteed by the simple fact that
spin operators of two different dimers commute.  However, because a
dimer can hold at most one triplon, the bosonic picture requires the
introduction of a hard-core constraint to exclude states with more
than one quasi--particle per dimer.  The constraint, which can be
interpreted as a strong short-range repulsion between the bosons,
poses a difficult theoretical problem.  But close to $\hcl$ their
density is small and collisions between bosons are rare; in this limit
interactions can be fully taken into account.  By particle-hole
symmetry, a similar simplification occurs near $\hcu$.

\section{BEC of triplons}

The nature of the ground state above $\hcl$, its interpretation as a
BEC of magnetic quasi-particles, and the tuning of the particle
density become particularly accessible if the spin Hamiltonian in
Equ. (\ref{eq:H}) is rewritten in the second-quantized form
\cite{Giamarchi,Nikuni}
\begin{equation}
\mathbf H = \sum_{i} (J_0-h)a^\dagger_i a_i
+\sum_{i,j} t_{ij}a^\dagger_i a_j + \frac{1}{2}
\sum_{i,j} U_{ij}a^\dagger_i a^\dagger_j a_j a_i
\label{eq:sq1}
\end{equation}
where $a^\dagger_i$ ($a_i$) create (annihilate) a boson on dimer $i$,
and $h=g\mu_B H$ is the effective field. The term $t_{ij}$ defined by
the transverse component of the {\em inter}-dimer interactions
$J^{x,y}_{1}$ etc describes hopping between sites $i$ and $j$ thus
endowing the triplons with kinetic energy, see Fig. 1b. $U_{ij}$ from
the longitudinal components $J^{z}_{1}$ etc is the repulsion energy
arising when two triplons occupy neighboring sites $i$ and $j$. The
density of triplons is directly controlled by the magnetic field that
acts thus as a chemical potential for the bosons \cite{shakespeare}.

Below $\hcl$ the ground state is a quantum-disordered paramagnet
formed by the singlet sea (triplon vacuum) and can be approximated by
the direct product of singlet states on each dimer, $|\psi \rangle_i =
|S,S^z \rangle_i$ with $S = S^z = 0$.  Once the spin gap is closed at
$\hcl$, a Bose condensate is formed.  Since the bottom of the triplon
band is located at a nonzero wavevector $\mathbf k_0 = (\frac{\pi}{a},
\frac{\pi}{a})$ (Fig. 1c), the wavefunction of the condensate varies
in space as $\exp{(i \mathbf k_0 \cdot \mathbf r)}$.  In this phase
the state of an individual dimer is well approximated by a coherent
superposition of the singlet and the $S^z = +1$ triplet: $|\psi
\rangle_i = \alpha_i(H) |0,0 \rangle_i + \beta_i(H) |1,1 \rangle_i$,
where the amplitudes $\alpha_i$ and $\beta_i$ depend on the magnetic
field $H$ \cite{Giamarchi,Matsumoto02,Matsumoto04}.

In the spin language, the condensate corresponds to a spontaneous
magnetic order formed by the transverse spin components $\langle S^x_i
\rangle$ and $\langle S^y_i \rangle$, violating the remaining
rotational $O(2)$ symmetry of the Hamiltonian (\ref{eq:H}).  To make
the analogy with the traditional BEC manifest, one can form a $U(1)$
order parameter $\langle S^x_i + i S^y_i\rangle$. The phase
corresponding to the angle of the spin in the $XY$ plane is thus the
phase of the wavefunction in the boson language.  At $\hcl$ the
paramagnetic phase at low fields makes a transition into a canted
antiferromagnet with long-range magnetic order in the plane
perpendicular to the field \cite{Giamarchi} (Figs. 1d and 3a).  The
staggering of the transverse components of magnetisation reflects a
nonzero wavevector of the condensate $\mathbf k_0$.  The critical
properties of the magnet in the vicinity of this phase transition are
governed by the quantum critical point (QCP) of the BEC universality
class located at $T=0$ and $H = \hcl$. Indeed close to $\hcl$ the
bosons are extremely diluted, and the effects of the interaction
become weak, despite its hard-core nature. Close to the QCP the phase
boundary $T_c(H)$ follows \cite{Giamarchi} a power law $T_c \propto
(H-\hcl)^\phi$ with a universal critical exponent $\phi=z/d$, which
depends only on the dimensionality ($d$) and dynamical critical
exponent $z=2$ for a quadratic triplon energy band. The upper critical
dimension for the QCP is $d_c=2$.

The BEC QPT has been observed in a growing number of dimer based
magnetic insulators, such as ACuCl$_3$ (A=Tl, K, NH$_4$), \bcso,
Cu(NO$_3$)$_2 \cdot$2.5D$_2$O, Cs$_3$Cr$_2$Br$_9$,
(CH$_3$)$_2$CHNH$_3$CuCl$_3$, and (C$_4$H$_{12}$N$_2$)Cu$_2$Cl$_6$
\cite{Nikuni,Ruegg03,Shiramura, Ruegg04, Matsumoto03,Jaime,Sebastian,Kramer,Grenier1,Grenier2,Garlea,Matsuda,Stone1,Stone2}.
The physics of triplon condensation
presented for the dimer system remains essentially unchanged since
these QCPs are in the same universality class. Quasi-one-dimensional
arrangements of $S=1$ moments in spin chains, e.g.  Haldane chains,
and even field-saturated frustrated antiferromagnets can be described
within the same framework, as was successfully done for nickel-based
materials and Cs$_2$CuCl$_4$, respectively
\cite{Zheludev,Zapf,Zvyagin,Coldea,Radu}.

Experimentally, the static and dynamic magnetic properties have been
studied at and around the BEC QCPs in such materials by many
experimental techniques.  The evolution of magnetic excitations across
the QCP at $\hcl$ in a three-dimensional network of dimers was studied
in TlCuCl$_3$, see Fig. 1a, 2a, and 2c
\cite{Matsumoto02,Matsumoto04,Ruegg03, Cavadini02}. The softening of
the triplet mode $S^z = +1$ in the quantum disordered phase, Fig. 2a,
is followed by a dramatic change in the nature of the excitation
spectrum above the critical field $\hcl$. In particular, as can be expected in a
system with a spontaneously broken XY symmetry in the plane perpendicular to the
applied magnetic field, a `Goldstone' mode, with a dispersion linear with momentum
appears \cite{Matsumoto02,Matsumoto04}.

Historically, the temperature dependence of the magnetisation $m_z(T)$
at fixed field $H>\hcl$ in TlCuCl$_3$ (Fig. 2d) and the scaling of the
critical temperature $T_c(H)$ were the key
experimental observations that confirmed the theoretical model
\cite{Giamarchi,Nikuni} of the magnon BEC in dimerized quantum
antiferromagnets. The minimum and cusp in $m_z(T)$ at the finite-temperature
transition can not be explained within simple mean-field theory
\cite{Tachiki}, but are a consequence of magnon condensation
\cite{Giamarchi,Nikuni} and as such also occur in lower-dimensional
magnets. For example comprehensive data from bulk experimental
measurements establish the phase diagram and critical exponents for
\bcso\/ (Fig. 2b).  The temperature crossover exponent is found to be
$\phi=2/3$, as expected for a BEC QCP in three dimensions.  However,
below 1 K and down to 35 mK the phase boundary unexpectedly becomes linear,
indicating that $\phi=1$ \cite{Sebastian,Batista,Vojta, Kramer}.  We discuss this
anomaly below. Even in the case where a full condensate does not exist,
such as in one-dimensional systems, similar minima in the magnetisation
can occur, but now as a simple crossover \cite{Oshikawa_BEC}.

\section{Comparison with other boson systems}

The correspondence between a Bose gas and an antiferromagnet is
summarized in Table~\ref{tbl:bec}.  Despite many similarities, there
are also a few important differences between triplons in a magnet and
atoms in a Bose gas
\cite{pitaevskii_becbook,Greiner,Bloch05,Bloch_RMP}.  Most
importantly, the number of atoms is usually controlled directly
(microcanonical ensemble), while the number of triplons is typically
set by the magnetic field acting like a chemical potential (canonical
ensemble).

Considerable differences also exist at the practical level. Triplons
are much lighter and have a much higher density than atomic gases.  As
a result, condensates survive to much higher temperatures: kelvins in
magnets, even room temperature in some cases \cite{Demokritov}, as
opposed to nanokelvins in atomic gases.  As solid-state systems, they
also allow for a variety of static and dynamic probes (magnetisation,
specific heat, NMR, neutrons, etc.). On the other hand, the cold
atomic systems allow potentially to study some out-of-equilibrium
situations that are hard to maintain in a solid-state device since the
condensate is still strongly coupled to the dissipative environment.

From the point of view of realization, the cold atomic systems have a
high degree of control and tunability in terms of the interactions and
structure. Indeed the structure of the lattice and the hopping of the
bosons can directly be controlled by varying the strength of an
optical lattice. The interaction can also be controlled via a Feshbach
resonance, allowing in principle to realize one's pet Hamiltonian.  In
spin systems the parameters can be changed only in a limited way by
the application of pressure or by changing the chemical composition.
Efforts in quantum chemistry produced good realizations of one-, two-,
and three-dimensional dimer materials, e.g.  TlCuCl$_{3}$, \bcso, and
(C$_{5}$H$_{12}$N)$_{2}$CuBr$_{4}$, respectively. In addition, by the very nature
of the mapping from spins to bosons, the spin systems offer a definite
advantage in reaching the limit of strong on-site repulsion as well as
the effects of interactions between nearest neighbors (see the next
section). The spin systems are therefore an optimal
starting point to study situations for which these ingredients are important.

The cold atomic systems have the important advantage that the phase
$U(1)$ symmetry is exact. In the magnets, the corresponding $O(2)$
symmetry in the plane perpendicular to the magnetic field can be
broken by weak anisotropic interactions (dipolar interactions,
crystalline anisotropies, spin-orbit coupling). Even when
symmetry-breaking terms are weak, they become important at low
temperatures and modify the physics in the vicinity of the QCP.  The
physics of BEC is also altered in the presence of coupling between
spins and lattice distortions. Fortunately,
it is possible to use our good control and knowledge of the pure BEC system
to treat theoretically these weak deviations as well
\cite{Misguich,Kolezhuk,Sirker,matsumoto_magnetostriction,orignac_magnetostriction}.
Finally the spin systems are excellent to study the critical behavior around
the quantum critical points, since the boson density can be finely controlled by the external
magnetic field, and the system is fully homogeneous. Intrinsic density inhomogeneities induced
by the confining trap in cold atomic systems, make a similar study more difficult.

\begin{table}[h]
\caption{\label{tbl:bec} Correspondence between a Bose gas and a quantum
antiferromagnet.}
\begin{center}

\begin{tabular}{|l|l|}
\hline Bose gas & Antiferromagnet\\ \hline \hline Particles & Spin
excitations ($S=1$ quasi-particles)\\ \hline Boson number $N$ & Spin
component $S^z$\\ \hline Charge conservation U(1) & Rotational
invariance O(2)\\ \hline Condensate wavefunction $\langle \psi(\mathbf
r) \rangle$ & Transverse magnetic order $\langle S^x_i + i
S^y_i\rangle$\\
\hline Chemical potential $\mu$ & Magnetic field $H$\\
\hline Superfluid density $\rho_s$ & Transverse spin stiffness\\
\hline Mott insulating state & magnetisation plateau\\ \hline
\end{tabular}
\end{center}
\end{table}

\section{Beyond simple BEC}
\label{sec:phases}

The interplay of an atomic lattice and strong boson interactions leads
to new phenomena beyond the simple BEC paradigm.  When the boson
density approaches a value commensurate with the lattice
periodicity, one often finds magnetisation plateaus where the
triplon density stays constant in a finite range of magnetic fields at
$T=0$.  We already discussed two simple examples of such plateaus for
$H< \hcl$ with zero magnetisation and for $H> \hcu$ where each site is
occupied by one triplon (Fig. 1d).  These simple $m_z=0$ or $1$
plateaus correspond to states with an integer number of bosons per
unit cell of the lattice. More generally, since in spin systems the
$S^z_{mi}S^z_{nj}$ terms translate into triplon repulsion,
magnetisation plateaus at fractional fillings can also occur
\cite{oshikawa_plateaus,cabra_plateaus}. Such intermediate plateaus
arise for strong enough repulsion between spin excitations on adjacent
lattice sites and beyond, and correspond in boson language to
Mott-insulator phases \cite{Rice,haldane_bosons,Fisher}. In the
simplest case (square lattice, repulsive nearest-neighbor
interactions) triplons may form an incompressible state by
`crystallizing' in a checkerboard-like pattern (Fig. 3c).  Unlike the
ground state of an integer magnetisation plateau (Fig. 3b), such a
state violates the translational symmetry of the lattice, and the
spontaneous order is characterized by an discrete order parameter
($Z_2$ in the case of the square lattice). Fractional plateaus require
strong magnon interactions (in comparison to the kinetic energy) and
are therefore less robust then integer ones.  Nonetheless, plateaus at
magnon fillings 1/3, 1/4, 1/8, and recently for other fractions
have been observed in \scbo\/ \cite{Kageyama,Kodama,Sebastian_cm}.  In
this material the geometrical frustration suppresses the kinetic
energy of magnons thus making magnon repulsion more pronounced
\cite{Miyahara}.

Outside the plateaus, magnons can be freely added to the condensate ensuring a
continuously varying magnetisation along the applied magnetic field. It has transverse
magnetic order violating the continuous $O(2)$ symmetry, is therefore gapless, and
corresponds to the superfluid phase of bosons. Its excitations are Goldstone modes,
characteristics of this superfluid phase \cite{Matsumoto02,Ruegg03}. The
incompressible plateau state and the gapless state with a continuously varying
magnetisation are again separated by a QCP.

While the dimensionality of an antiferromagnet is determined mostly by
quantum chemistry, it can also change on the fly as the magnet is
cooled down.  The ''Han purple'' \bcso\ contains layers of dimers with
substantial exchange couplings between the layers.  Yet the critical
behavior exhibits a surprising crossover from three-dimensional at
high temperatures to two-dimensional close to the QCP 
\cite{Sebastian,Kramer}. Two effects contribute to this dimensional reduction
at low temperatures. The interplane triplon
hopping is suppressed by geometrical frustration: it vanishes exactly
at the wavevector of the condensate \cite{Sebastian,Batista}.
Additionally, there are inequivalent layers with different values of
the spin gap \cite{Ruegg07,Kramer,Vojta}.

Another spectacular example of reduced dimensionality is provided by
(C$_{5}$H$_{12}$N)$_{2}$CuBr$_{4}$ \cite{Watson,Lorenz} and related
organo--metallic compounds. These materials can be thought of as a
collection of chains of dimers nearly decoupled from each other. As a
result, a large part of the $H-T$ phase diagram can be understood by
starting in the one-dimensional limit
\cite{Giamarchi,chitra_spinchains_field,mila_field,furusaki_dynamical_ladder,usami_ladder_dmrg,hikihara_ladder_dmrg}.  At low energies the
triplons on a single chain form a Luttinger liquid
\cite{giamarchi_book_1d}.  Such a compound is thus particularly suited
to address the interesting question of the dimensional crossover
between a one-- and three--dimensional system. In the former, the
spins will more behave as fermions, as is well known from the
Jordan-Wigner transformation, and in the latter, at temperatures and
fields for which residual exchange coupling between the ladders
becomes important, the spins are much more behaving as ``regular
bosons'' \cite{orignac_NMR_BEC}. The excellent degree of control over
the density by the magnetic field as well as the various probes
providing access to the dynamical spin-spin correlations make
quasi-one-dimensional magnets ideal for studying the dimensional
crossover, a phenomenon of general importance for several major areas
in physics \cite{giamarchi_book_1d}.

We would like to mention the tantalizing possibility of finding an
exotic phase known as the supersolid \cite{Prokofev}.  A hybrid
of a solid and a superfluid, this phase violates both the
translational symmetry by forming a density wave and the $U(1)$ phase
symmetry by exhibiting transverse magnetic order (Figs. 3b-c).  It has
been conjectured long ago \cite{Andreev,Chester,Leggett} and its
existence is under intense debate at the
moment following experiments in helium \cite{Chan,balibar}.  Such a phase
could occur at the end of a fractional plateau because the transition
must accomplish two tasks: create a superfluid and destroy a solid
(i.e. restore the broken translational symmetry).  According to
Landau's theory of phase transitions, such transitions are either
discontinuous or happen in two stages. Several theoretical
calculations show that supersolids exist in realistic magnetic models
\cite{Troyer,Heidarian,Melko,Ng,Sengupta,Laflorencie}. The observation
of such a phase would thus be an interesting realization of the
physics of interacting bosons on a lattice.

Last but not least, advances in growth techniques may allow in the
future to study the effects of disorder on bosons \cite{giamarchi_loc,Fisher,Nohadani_BG}
and the Bose glass by creating bond disorder, and the influence of impurities on the
quantum-critical states \cite{Mikeska,Xu_hole}.  Other directions
include the thermodynamics of strongly interacting bosons
\cite{Xu,Ruegg05} and quantum coherence at the mesoscopic scale
\cite{Xu_T}.

\section{Conclusions and Outlook}
\label{sec:outlook}

Quantum spins and quantum dimer systems offer remarkable opportunities
to study phenomena related to the Bose-Einstein condensation of
interacting quantum particles.  The high density and low mass of spin
excitations, magnons or triplons, leads to a robust nature of the
condensate, which survives to temperatures as high as 10 K.  The
availability of numerous model magnets and physical probes has enabled
a detailed study of the critical phenomena and magnetic properties in
the vicinity of the BEC quantum phase transition. The spin systems
have thus proven to be nicely complementary to other systems such as
the cold atomic gases for the investigation of BEC.

In addition to providing a direct connection with the remarkable BEC
phenomenon, the boson picture offers an intuitive understanding for
complex quantum properties of matter that are much less physically
transparent in the original spin language, and provides access to a
cornucopia of exciting new problems and questions. For example, a
classical description as canted antiferromagnets of the dimerized
magnetic materials discussed here in the field-induced order phase
misses up to 100\% of the relevant spin physics close to the intrinsic
quantum phase transition. Here the possibility to fine tune the
density of bosons by using the magnetic field provides a tool to
determine the role of the interactions in such systems and a wide
range of lattice geometries and dimensionalities is available for
study. In combination with the large number of experimental probes for
spins this allows to challenge our knowledge of exotic phases of
strongly interacting quantum particles, such as BEC in various
dimensions, Luttinger-liquid physics, commensurate solids with a
fractional number of bosons per unit cell, and supersolids combining
superfluidity with a broken translational symmetry.

\vskip 0.5cm
\noindent {\bf Acknowledgments}

This work was supported in part by the Swiss NSF under NCCR MaNEP, a Wolfson Royal Society Research Merit Award, and the US National Science Foundation.

\vskip 0.5cm
\noindent {\bf Competing financial interests}

The authors declare no competing financial insterests.

\vskip 0.5cm

\noindent Correspondence to T.G., thierry.giamarchi@physics.unige.ch, C.R., c.ruegg@ucl.ac.uk, or O.T., olegt@jhu.edu.

\newpage

\begin{figure}[h]
\includegraphics[width=0.9\textwidth]{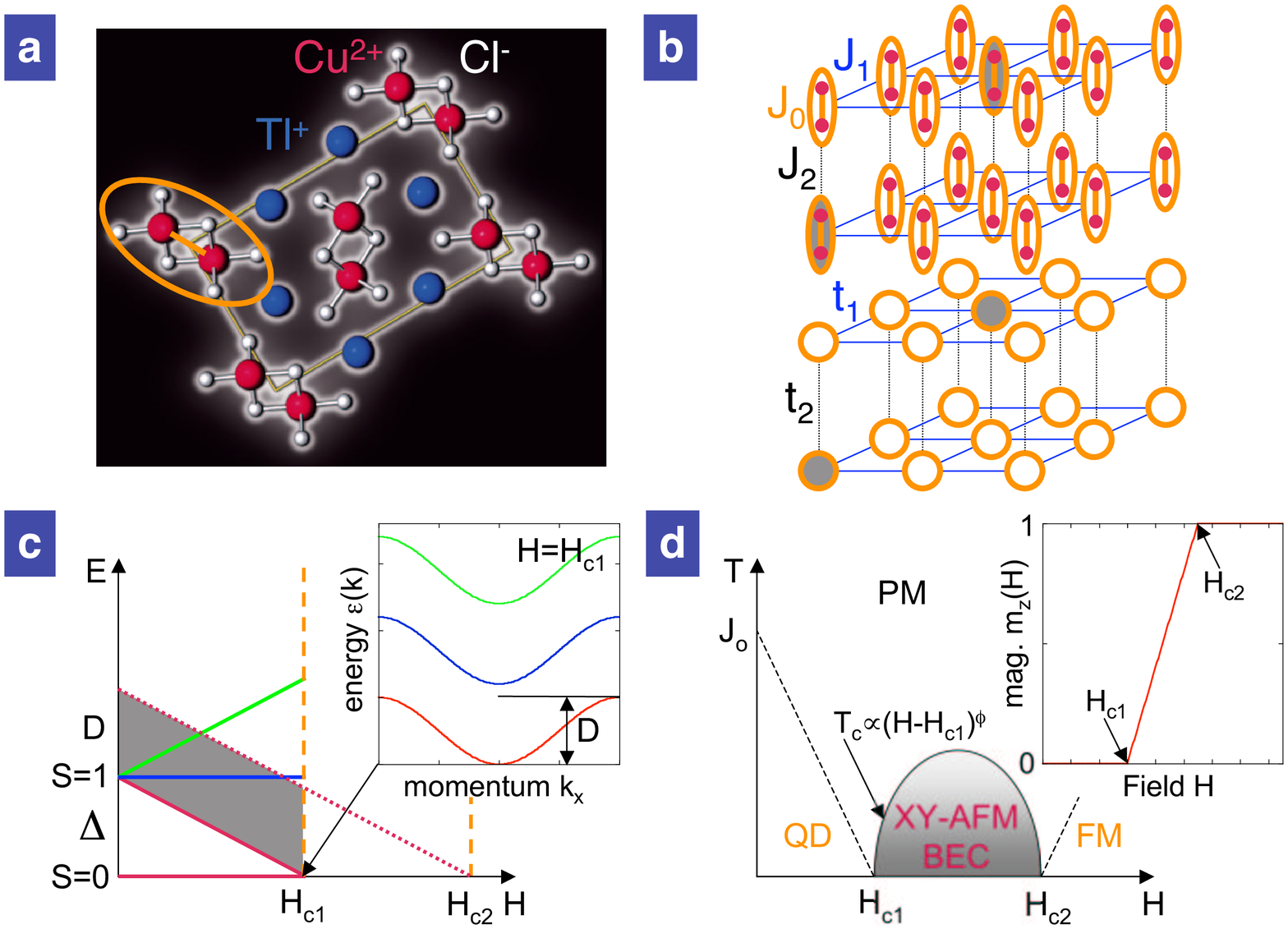}
\caption{BEC of magnons in dimerized quantum antiferromagnets: (a)
  Dimers in the real material TlCuCl$_3$ with $S=1/2$ from Cu$^{2+}$
  ions and superexchange via Cl$^-$
  \cite{Nikuni,Ruegg03,Cavadini01,Cavadini02,Matsumoto02,Matsumoto04,
    Ruegg05}.
  (b) Dimers on a square lattice with dominant antiferromagnetic
  intradimer interaction $J_0$ and interdimer interactions
  $J_i$. Triplet states (gray, top) are mapped on quasi-particle
  bosons (triplons, bottom). (c) Zeeman splitting of the triplet modes
  with gap $\Delta$ and bandwidth $D$ at $\mathbf k_0 =
  (\frac{\pi}{a}, \frac{\pi}{a})$. Dispersion of triplons at the
  critical field $\hcl$
  \cite{Matsumoto02,Matsumoto04,Ruegg03,Cavadini01,Cavadini02}. (d)
  Resulting phase diagram with paramagnetic (PM), quantum disordered
  (QD), field-aligned ferromagnetic (FM), and canted-antiferromagnetic
  (XY-AFM) phase, where BEC of magnons occurs.  Close to $H_{c1}$ and
  $H_{c2}$ the phase boundary follows a power--law $T_c \propto
  (H-\hcl)^\phi$ with a universal exponent $\phi=2/3$ for BEC of
  magnons \cite{Giamarchi,Nikuni}.  Magnetisation curve $m_z(H)$ for 3D
  dimer spin system with plateau at $m_z=0,1$ (gapped)
  \cite{Nikuni,Jaime,Sebastian}.}
\label{fig1}
\end{figure}

\newpage

\begin{figure}[h]
  \includegraphics[width=0.9\textwidth]{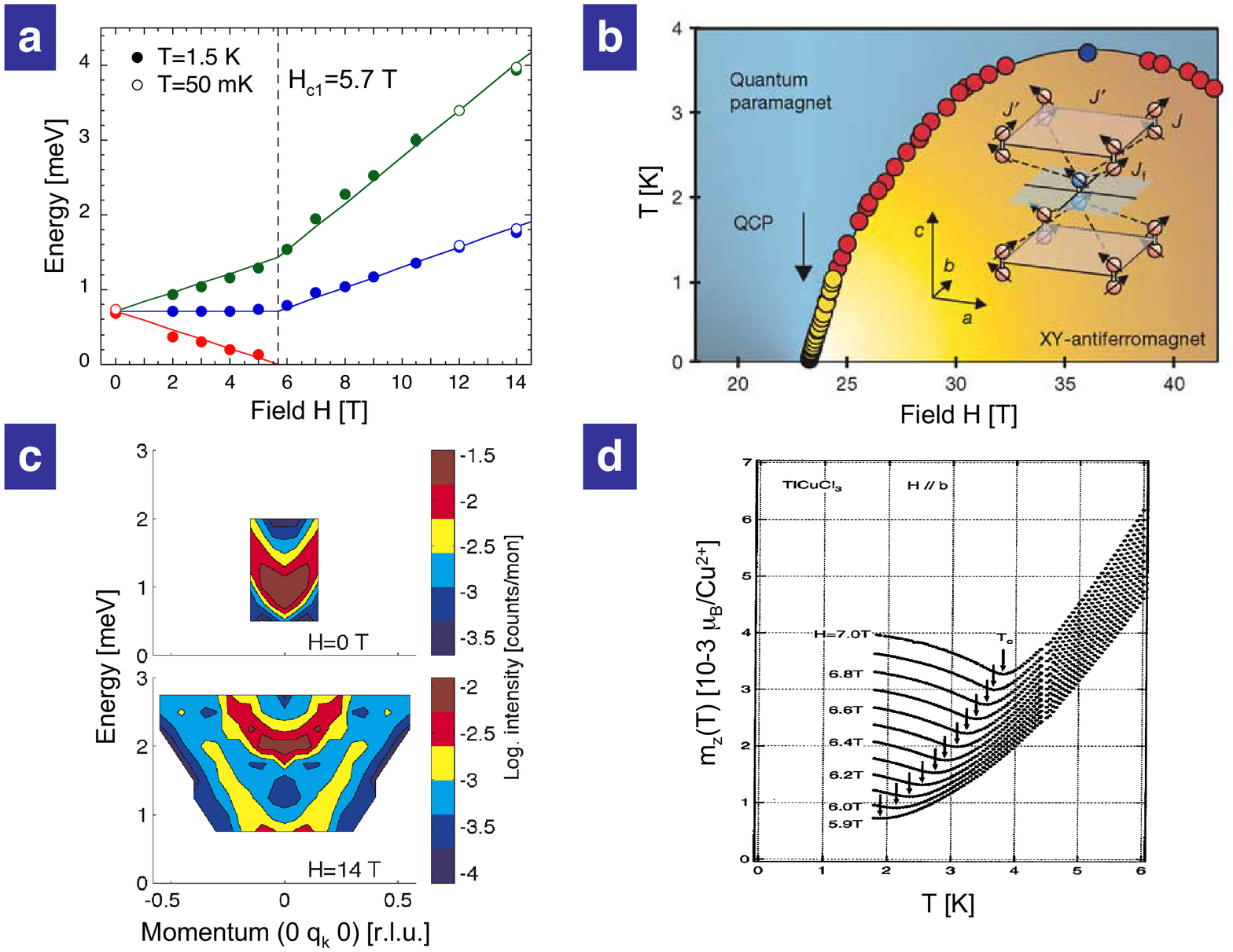}
  \caption{Experimental results on magnon BEC: (a) Zeeman splitting of
    the triplet modes in TlCuCl$_3$ up to $H>\hcl$ measured by
    inelastic neutron scattering \cite{Ruegg03,Glazkov,Kolezhuk}. (b)
    Phase diagram of \bcso~measured by torque magnetisation,
    magneto-caloric effect, and specific heat \cite{Jaime,Sebastian}.
    Dimensional reduction was reported in this material with a
    crossover from the 3D-BEC critical exponent $\phi=2/3$ to $\phi=1$
    for 2D at temperatures close to the QCP \cite{Sebastian}. (c)
    Excitations in the BEC of triplons realized in TlCuCl$_3$
    \cite{Ruegg03,Glazkov,Kolezhuk}. Goldstone mode with linear
    dispersion around $\mathbf k_0$. Spin anisotropy generally leads
    to a spin gap in real materials
    \cite{Glazkov,Kolezhuk,Sirker}. (d) Temperature-dependence of the
    magnetisation $m_z(T)$ in TlCuCl$_3$ for fixed magnetic field $H$,
    as indicated \cite{Nikuni}.  Minima at the finite-temperature
    phase transition (vertical arrows), as expected for a BEC of
    triplons.}
\label{fig2}
\end{figure}

\newpage

\begin{figure}[h]
\includegraphics[width=0.9\textwidth]{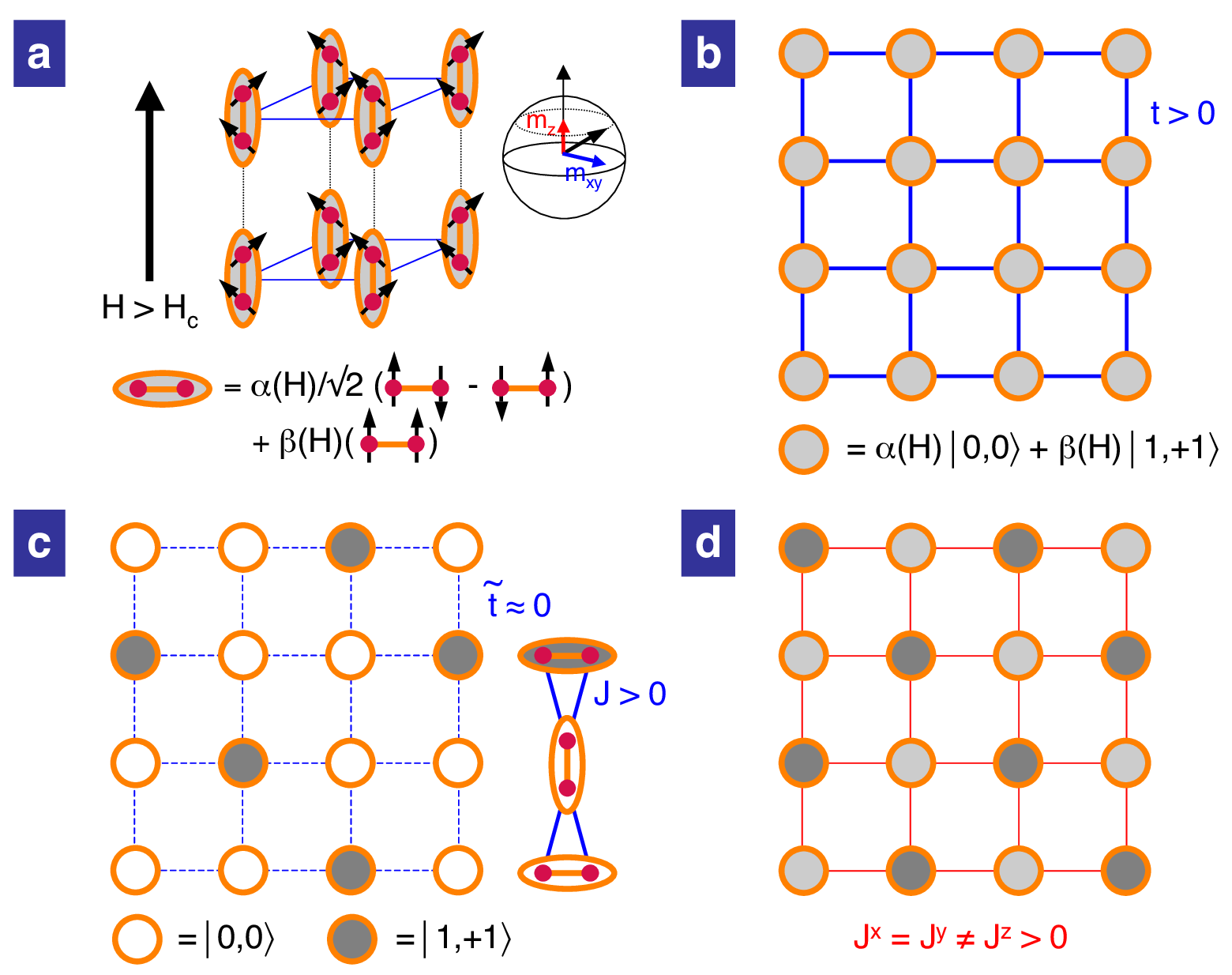}
\caption{Lattice boson phases: (a) Canted antiferromagnetic state at
  $\hcl<H<\hcu$ in a BEC of magnons. Longitudinal $m_z$, see Fig. 1d,
  and transverse ($XY$) magnetisation $m_{xy}$. (b) Triplon condensate
  in superfluid phase. The hopping amplitude $t$ of the
  quasi-particles leads to a uniform mixture of singlet and triplet
  states. (c) Analog of the Mott-insulating phase for triplons with
  magnetisation plateau at $m_z=1/3$: triplons ''crystallize'' to form
  a superstructure. This phase is partially observed in \scbo, where
  quasi-particle hopping is strongly suppressed by geometrical
  frustration \cite{Kageyama, Kodama,Miyahara,Rice}. (d) Supersolid on
  a square-lattice of dimers with exchange anisotropy as proposed
  theoretically \cite{Ng, Laflorencie}, and for other lattice
  geometries \cite{Troyer,Heidarian,Melko,Sengupta}. Both translation
  and $U(1)$ symmetry are spontaneously broken, i.e. coexistence of
  superfluid and ''crystal'' of lattice bosons. }
\label{fig3}
\end{figure}


\begin{thebibliography}{99}
\bibitem{shakespeare} Some purists object to the use of ``chemical
  potential'' for the magnetic field term. Note that here the
  total component of $S^z$ commutes with the Hamiltonian so that the
  total magnetisation along $z$ and thus the total number of bosons is
  conserved. We also refer those purists to Shakespeare, Romeo and
  Juliet II.2.43: ''What's in a name? That which we call a rose By any
  other word would smell as sweet''.  We thus stick to the terminology
  chemical potential in the following.
\end{thebibliography}
\end{document}